\documentclass[a4paper,UKenglish]{oasics}

\usepackage{microtype}
\usepackage{xspace}
\makeatletter
\newcommand{\mezzo}{%
  \texorpdfstring{%
    \protect\mbox{%
      \protect\textit{\protect\textsf{%
        Me\protect\raisebox{-.1ex}z\protect\hspace{-.3ex}{\protect\raisebox{.2ex}z}\protect\hspace{-.1ex}o%
      }}%
    }%
  }{Mezzo}\xspace
}

\@ifpackageloaded{listings}{
  \lstset{
    language=caml,
    morekeywords={data,val,consumes,adopts,take,give,from},
  }
  \def\li{\lstinline}
  \def\sqli|#1|{``\texttt{\small #1}''}
}

\makeatother

\newcommand{\kw}[1]{\ensuremath{\text{\textsf{#1}}}}

\newcommand{\kleft}{\kw{left}}
\newcommand{\kright}{\kw{right}}

\makeatletter
\@ifundefined{mode}{
  
}{}
\makeatother

\newcommand{\tyarrow}[2]              {\ensuremath{#1 \rightarrow #2}}

\newcommand{\tyconcrete}[2]   {\ensuremath{\kw{#1}\,\{ #2 \}}}

\newcommand{\tysingleton}[1]{\raisebox{0.0mm}{\mbox{\kw{=}}}#1}

\newcommand{\tystar}[2]               {\ensuremath{#1 \ast #2}}
\newcommand{\tyatomic}[2] {\ensuremath{#1\,@\ #2}}

\newcommand{\tyint}                   {\kw{int}}
\newcommand{\tystring}                {\kw{string}}

\newcommand{\tylist}[1]               {\kw{list}\;#1}

\newcommand{\tymtree}[1]              {\kw{mtree}\;#1}

\title{Illustrating the \mezzo programming language}

\author[1]{Jonathan Protzenko}
\affil[1]{INRIA\\
  Rocquencourt, France\\
  \texttt{jonathan.protzenko@ens-lyon.org}}
\authorrunning{J. Protzenko}

\Copyright{J. Protzenko}


\serieslogo{}
\volumeinfo
  {Christine Choppy and Jun Sun}
  {2}
  {1st French Singaporean Workshop on Formal Methods and Applications 2013 (FSFMA'13)}
  {31}
  {1}
  {1}
\EventShortName{FSFMA'13}
\DOI{10.4230/OASIcs.FSFMA.2013.<firstpage>}

\begin{document}

\maketitle

\begin{abstract}
  When programmers want to prove strong program invariants, they are usually
  faced with a choice between using theorem provers and using traditional
  programming languages. The former requires them to provide program proofs,
  which, for many applications, is considered a heavy burden. The latter provides
  less guarantees and the programmer usually has to write run-time assertions
  to compensate for the lack of suitable invariants expressible in the
  type system.

  We introduce \mezzo, a programming language in the tradition of ML, in which the
  usual concept of a type is replaced by a more precise notion of a permission.
  Programs written in \mezzo usually enjoy stronger guarantees than programs
  written in pure ML. However, because \mezzo is based on a type system, the
  reasoning requires no user input. In this paper, we illustrate the key
  concepts of \mezzo, highlighting the static guarantees our language provides.
\end{abstract}

\section{Introduction}

Type systems help programmers reason about the types of the manipulated objects,
which embed information about their memory structure. Programs which obey a
strong static discipline, such as that of ML, therefore have the powerful
property that that they cannot go wrong. In other words, a well-typed program
will not lead to a segmentation fault.

In practice, programmers want to reason beyond the memory layout of objects.
Real-world objects often follow a protocol, going through different
\emph{states}, that only permit certain operations. A file descriptor starts
\emph{uninitialized}, then it may move to \emph{ready}, before being
\emph{closed}. The ``open'' operation may only be performed on an uninitialized
file descriptor, while the ``close'' operation only works when the file
descriptor is ready. Thus, the file descriptor changes \emph{states}, while
preserving its \emph{type}. However, traditional type systems fail to help
programmers statically check \emph{state} invariants.

\mezzo~\cite{mezzo} is a programming language that reads and feels like ML, but
that is equipped with a more powerful type system, which attempts to answer the
above concerns.  Since \mezzo has a more rigid typing discipline than ML, some
programs that previously type-checked in ML will be deemed too unsafe.
Conversely, as \mezzo allows more precise reasoning, some programs that
previously could not be type-checked in ML will be understood.

In \mezzo, the notion of state and that of a type are conflated. An object which
moves from a state to another is an object whose type changes. For instance, the
``open'' operation will change the type of its argument from
\emph{uninitialized} to \emph{ready}. This design choice requires careful
reasoning about \emph{ownership}. Indeed, it is crucial that no other part of
the system sees the object with its previous type, as this would naturally lead
to an inconsistency, and protocol violations.  Therefore, the type
system should track ownership and avoid undesired
aliases, as having two distinct names for the same object makes it difficult to
ensure consistency.

Literature offers a wealth of related work, and \mezzo draws inspiration from
several areas.
The biggest source of inspiration is \emph{Separation
Logic}~\cite{reynolds-02,ohearn-07}, a program logic that describes the state of
the heap. In separation logic, asserting that an object has a given type amounts
to owning that object. We reuse that principle, but turn it into a type system
through our notion of permission, while also extending the reasoning to
non-mutable portions of the heap.
The \emph{Plaid Project}~\cite{bierhoff-aldrich-07} annotates references to
objects with permission. This asserts both what one is allowed to do with the
object, as well as what others may do with it. Our permission mechanism supports
similar reasoning, but unifies both the state and the mutation invariants of an
object, using permissions.
In \mezzo, pointers can be copied, while the original permission on the object
remains. We keep track of aliasing through the use of singleton types, inspired
by \emph{Alias Types}~\cite{alias-types-00}.
Our notion of affinity, expressing that items may be used at most once,
while others may be used freely, is reminiscent of \emph{Linear
Types}~\cite{morrisett-al-07}.

We begin with an introduction to permissions, a core concept in \mezzo.  Next,
we discuss a case study, emphasizing several possible mistakes ruled out by our
typing discipline. Finally, we give an overview of other \mezzo features and
conclude with pointers to reference material.

\section{An introduction to permissions}

The central concept in \mezzo is that of a \emph{permission}. While in the
$\lambda$-calculus we say that ``$x$ has type $t$'', in \mezzo we say we have
permission $\tyatomic xt$, which we read ``$x$ at type $t$''. We
can think of a permission as a token that grants access to variable $x$
with type $t$.

Unlike traditional typing judgements, permissions are transient. The user may
possess $\tyatomic xt$ at some point of the program, and have instead $\tyatomic
xu$ later on, which accounts for the fact that the type of $x$ changed from $t$
to $u$, i.e. that the \emph{state} of $x$ changed.

A permission may be obtained by creating a new value. Writing
\sqli|let x = (1, "hello")| yields $\tyatomic{x}{(\tyint,\kw{string})}$,
granting its owner the right to use $x$ as a tuple of an $\kw{int}$ and a string.

\subsection{Permissions control effects}

Permissions appear in the signature of functions. Let us consider the type of
the \li|length| function, which, as the name implies, computes the length of a
list. The square brackets stand for universal quantification.
\begin{lstlisting}
val length: [a] (x: list a) -> int
\end{lstlisting}
The introduction of the
name $x$, along with its type $\tylist a$, is syntactic sugar: the function
expects an argument named $x$, along with a permission $\tyatomic{x}{\tylist
a}$. Conceptually, the function demands a token of ownership from its caller, so
as to iterate over the list and compute its length.  Hence, whenever one wishes
to \emph{call} the function on argument $x$, a permission $\tyatomic{x}{\tylist
a}$ will be removed from the caller's current set of permissions.

Unless otherwise specified, and as a syntactic convention, we understand the
permission $\tyatomic{x}{\tylist a}$ to be returned to the caller. Therefore,
after the function call, the caller will regain $\tyatomic{x}{\tylist a}$, along
with a permission $\tyatomic r\tyint$, where $r$ is the name of the return
value.

Another, more sophisticated function type, is that of the \li|annotate|
function. It takes a \emph{mutable} binary search tree of strings and modifies
each node to store a pair, consisting of its original value and the size of the
corresponding subtree, which the function returns.
\begin{lstlisting}
val annotate: (consumes t: mtree string)->(int|t@ mtree (string,int))
\end{lstlisting}
Thanks to the \li|consumes| keyword, this function now takes a permission
$\tyatomic{t}{\tymtree\tystring}$ from the caller and returns a \emph{different}
permission for $t$, namely $\tyatomic{t}{\tymtree{(\tystring, \tyint)}}$.
Therefore, the type of $t$ changes through a call to \li|annotate|.  This is a
\emph{type-changing update}, which the permissions mechanism accurately
describes.

\subsection{Permission denote ownership}

In order for the above function to be sound, no one else must own a copy of
$\tyatomic{t}{\tymtree\tystring}$, since that copy would be invalidated after
calling \li|annotate|. Therefore, permissions that denote mutable portions of
the heap must be uniquely owned. We say that the permission
$\tyatomic{t}{\tymtree\tystring}$ is \emph{exclusive}. The type system enforces
this policy, by preventing exclusive permissions from being duplicated.

Conversely, $\tyatomic{x}{(\tyint,\tystring)}$ denotes read-only knowledge, as
our tuples, integers and strings are immutable. This knowledge is permanent, as
the type of $x$ will never change. Hence, it is sound to share that
information. We say that the permission is \emph{duplicable}, and
the type-checker will allow the user to obtain as many copies of the permission
as desired.

A permission $\tyatomic xt$ therefore states not only that ``$x$ has type $t$'',
but also that ``we own $x$ at type $t$''. The user (and the type-checker) can,
by looking at $t$, infer whether $t$ is exclusive or duplicable, i.e.  whether
they have an exclusive, read-write access, or a shared, read-only access to the
object. The details for this procedure, called \emph{mode inference}, are
available~\cite{icfp-long}.

Some permissions are neither exclusive nor duplicable; they are said to be
\emph{affine}. Such a permission is $\tyatomic xa$, where $a$ is an abstract
type variable, which may occur in the body of a function polymorphic in $a$. We
have to be conservative and make no assumptions on $a$.

\subsection{Permissions track aliasing}

At any given program point, a current permission is available, granting us
ownership of a part of the heap. Combining \emph{atomic} permissions
of the form $\tyatomic xt$ into a composite permission is achieved
using the $\ast$ connective; we say that the conjunction of $p$ and $q$ is
$\tystar pq$. This conjunction is reminiscent of separation logic. Indeed, if
$t$ and $t'$ are both exclusive, the conjunction $\tystar{\tyatomic
xt}{\tyatomic y{t'}}$ implies that, because one cannot hold two exclusive
permissions for the same variable, $x$ and $y$ must be distinct.  This is a
\emph{must-not-alias constraint} and we state that our $\ast$ conjunction is
separating on exclusive portions of the heap.

Moreover, $\ast$ extends the conjunction of separation logic to
non-exclusive portions of the heap. If $t'$ is duplicable, then
$\tystar{\tyatomic xt}{\tyatomic y{t'}}$ yields no information: $x$ and $y$ may
or may not be aliases, and the conjunction just has to be consistent. The same
situation holds if both $t$ and $t'$ are duplicable.
Normally, conjunctions are consistent: if $\kw{Nil}$ denotes the empty list
cell,
$\tystar{\tyatomic x{\tylist\tyint}}{\tyatomic x\kw{Nil}}$ is a
conjunction that is always consistent. However, inconsistent conjunctions
exist, such as
$\tystar{\tyatomic x{\tymtree\tyint}}{\tyatomic x{\tymtree\tyint}}$.
Our system has been proved sound, meaning that a program cannot reach a
configuration where this conjunction holds. This point in the program
is unreachable: it is statically ruled out as ``dead code''.

These must-not-alias constraints, expressed implicitly in a conjunction, are
completed by \emph{must-alias constraints}, which are expressed using
\emph{singleton types}. A singleton type is of the form $\tysingleton y$, where
$y$ is a program variable. This type has exactly one inhabitant: $y$ itself.
Therefore, having $\tyatomic x{\tysingleton y}$ means that $x$ and $y$ are
actually equal; in particular, if they are pointers, they point to the same
value. We write this using syntactic sugar as $x = y$.

A singleton type appears whenever one creates an alias. If $\tyatomic xt$ holds,
then writing \li|let y = x in ...| yields a new permission $x = y$, without
duplicating the original permission on $x$, which greatly simplifies reasoning.
Singleton types are pervasive in \mezzo; they are particularly useful when
combined with \emph{structural types}.

\begin{lstlisting}[caption={Definition of mutable, binary
trees},captionpos=t,label={lst:mtree}]
data mutable mtree a =
  | Null
  | Node { left: mtree a; value: a; right: mtree a }
\end{lstlisting}

Listing \ref{lst:mtree} above shows how to define an algebraic data type in
\mezzo. Defining such a type allows one to obtain permissions of the form
$\tyatomic x{\tymtree a}$. However, the type of $x$ may be refined using a
$\kw{match}$ expression; one may trade this permission for a more precise one,
of the form $\tyatomic x{\tyconcrete{Node}{ \kleft: \tymtree a; \kw{value}: a;
\kright: \tymtree a }}$. To understand what it means to own a value with such a
type, let us rewrite this compact permission, introducing names for the three
fields of $x$, as:
$\tystar
{\tyatomic x{
  \tyconcrete{Node}{ \kleft: \tysingleton l; \kw{value}: \tysingleton v; \kright: \tysingleton r }
}}
\tystar
{\tyatomic l{\tymtree a}}
\tystar
{\tyatomic va}
{\tyatomic r{\tymtree a}}
$.

The ownership semantics of the compact permission can now be understood as
stating that we own a memory block at address $x$ of size four, containing a tag
$\kw{Node}$ and three fields. We also own two mutable trees located at addresses
$l$ and $r$, along with a value of type $a$ named $v$. The points-to
relationships are represented by the singleton types.
Similarly, the meaning of a nominal permission, such as $\tyatomic x{\tymtree a}$, is
the disjunction of the meaning of its unfoldings $\tyatomic x{\kw{Null}}$ and
$\tyatomic
x{\tyconcrete{Node}{ \kleft: \tymtree a; \kw{value}: a; \kright: \tymtree a
}}$.

\section{A \mezzo case study}

We now discuss a motivating example that, by using all the mechanisms described
above, avoids several pitfalls. This example, as shown in listing \ref{lst:bst},
consists of splitting a mutable binary search tree. The function \li|split|
``steals'' the ownership of its argument $t$ from its caller, and returns two
binary search trees: the first one containing all values $v \leq k$ and the
second one containing all values $v > k$.  The \li|split| function abstracts
over the comparison function \li|cmp|. We omit the re-balancing of the tree and
focus on a self-contained example.

There are several pitfalls that await the programmer when writing such code.
The user may inadvertently copy a key: this would violate the invariant that the
two sub-trees form a partition.  The user may return, as the right sub-tree, a
pointer into the left sub-tree: this would create undesired sharing, leading to
subtle bugs when two concurrent threads will want to access the two sub-trees,
assuming that they are distinct. The permissions mechanism, by ensuring that
exclusive knowledge cannot be duplicated, enforces these invariants
statically.

\begin{lstlisting}[
  caption={In-place splitting of a mutable binary search tree},
  captionpos=t,
  label={lst:bst},
  %float,abovecaptionskip=-\medskipamount,
  numbers=left,
  numberstyle=\footnotesize\ttfamily\itshape,
  numbersep=2ex,
  language=Caml
]
val rec split_right [a] (
  consumes parent: Node { left: mtree a; value: a; right = child},
  consumes child: mtree a,
  k: a,
  cmp: (a, a) -> int
): (mtree a | parent @ mtree a) =
  match child with
  | Null -> Null
  | Node ->
      if cmp (child.value, k) <= 0 then
        split_right (child, child.right, k, cmp)
      else begin
        let left_leq, left_gt = split (child.left, k, cmp) in
        parent.right <- left_leq;
        child.left <- left_gt;
        child
      end end

and split [a] (consumes t: mtree a, k: a, cmp: (a, a) -> int)
  : (mtree a, mtree a) =
  match t with
  | Null -> Null, Null
  | Node ->
      if cmp (t.value, k) <= 0 then begin
        let right_gt = split_right (t, t.right, k, cmp) in
        t, right_gt
      end else begin
        let left_leq, left_gt = split (t.left, k, cmp) in
        t.left <- left_gt;
        left_leq, t
      end end
\end{lstlisting}

\subsection{The \li|split| function}

The \li|split| function is the entry point. At the start of the
function body, the permission is:
$$
\tystar
{\tyatomic t{\tymtree a}}
\tystar
{\tyatomic ka}
{\tyatomic {\kw{cmp}}{\tyarrow{(a, a)}{\tyint}}}
$$
The function begins by matching on its argument $t$. If $t$ is $\kw{Null}$, two
empty trees are returned. In the converse case, the permission on $t$ is refined
to:
$$\tystar
{\tyatomic t{
  \tyconcrete{Node}{ \kleft: \tysingleton l; \kw{value}: \tysingleton v; \kright: \tysingleton r }
}}
\tystar
{\tyatomic l{\tymtree a}}
\tystar
{\tyatomic va}
{\tyatomic r{\tymtree a}}
$$
In the case that $t.\kw{value} \leq k$ holds, a call to \li|split_right|, whose
meaning we will explain in the next section, is made. In the case that $k <
t.\kw{value}$ (line 29), values greater than $k$ may be found in the left
sub-tree.  The recursive call yields a partition of left sub-tree, consuming
$
{\tyatomic l{\tymtree a}}
$,
while producing
$
\tystar
{\tyatomic{\kw{left\_leq}}{\tymtree a}}
{\tyatomic{\kw{left\_gt}}{\tymtree a}}
$.
We re-attach values greater than $k$ into $t.\kleft$, thus changing the
permission of $t$ into
$
\tyatomic t{
  \tyconcrete{Node}{ \kleft: \tysingleton \kw{left\_gt}; \kw{value}: \tysingleton v; \kright: \tysingleton r }
}
$
Values lesser or equal to $k$ are returned, along with $t$, which now contains
the set of all possible values greater than $k$.

Which mistakes could an absent-minded programmer do? A first one would be
forgetting to perform the assignment \sqli|t.left <- left\_gt|, at line 30. The
permission for $t$ would still be 
$
\tyatomic t{
  \tyconcrete{Node}{ \kleft: \tysingleton l; \kw{value}: \tysingleton v; \kright: \tysingleton r }
}
$. However, the call to \li|split|, at line 29, consumed the permission for $l$:
it is no longer available, thus preventing this type from being folded back to
$\tymtree a$, when exiting the function.  \mezzo would reject this program.

Another beginner's mistake would be to return the value \li|(left_gt, t)|.
Following the return type \li|(mtree a, mtree a)|, \mezzo would consume
${\tyatomic{\kw{left\_gt}}{\tymtree a}}$ to prove that \li|left_gt| is a tree.
Next, \mezzo would have to prove that \li|t| is a tree, using permission
$
\tyatomic t{
  \tyconcrete{Node}{ \kleft: \tysingleton \kw{left\_gt}; \kw{value}: \tysingleton v; \kright: \tysingleton r }
}
$. Specifically, this implies proving that \li|t.left|, also known as
\li|left_gt|, is also a tree. Unfortunately, that exclusive permission was
already consumed. This situation is therefore rejected.

\subsection{The \li|split_right| function}

The \li|split_right| function is written in a different style, as it takes a
non-null \li|parent| tree, along with its right \li|child|. After the function
call, the parent is still a tree, holding all values lesser or equal to $k$,
while the returned tree contains all values greater than $k$.

The call to \li|split_right| at line 11 is legal, as we know that \li|child| is
a \li|Node|, which justifies using it as the first argument.  \mezzo statically
checks that the second argument is indeed the right child of the first: this
information is known statically, due to the use of singleton types. This
contrasts with the usage in traditional imperative languages, where typical code
would rely on a loop and two mutable variables, with the implicit invariant that
one is the right child of the other. Here, the invariant is made explicit and
\mezzo can rule out misuses.

The type-checker applies recursive reasoning. After the call to \li|split_right|
at line 11, if $\kw{ret}$ denotes the return value of the recursive call, the
remaining permission is:
$$
\tystar
{\tyatomic{\kw{parent}}{\tyconcrete{\kw{Node}}{
  \kleft: \tymtree a; \kw{value}: a; \kright: \tysingleton{\kw{child}}
}}}
\tystar
{\tyatomic{\kw{child}}{\tymtree a}}
{\tyatomic{\kw{ret}}{\tymtree a}}
$$
The type-checker then performs one last folding, to obtain the desired return
type for the function. Note that the function call is tail-recursive, while the
reasoning is \emph{not}. Indeed, the use of recursive functions with distinct
pre- and post-conditions yields more expressiveness than the use of traditional
loops. This allows for stronger invariants.

\section{Conclusion}

Due to its \emph{permissions} formalism, the \mezzo language manages to state
precise invariants for programs that rely on mutable state, thus preventing
several programming mistakes. The key mechanisms enforcing this rely on
ownership, linearity and singleton types.

Permissions, as presented here, cannot account for non tree-shaped aliasing
patterns. However, \mezzo offers several mechanisms for evading this
restriction, e.g. locks, Boyland's nesting~\cite{boyland-nesting-10} and our own
adoption/abandon mechanism.  A more thorough discussion can be
found~\cite{icfp-long}, which details the language with typing rules and a
formal definition of permissions. A gallery of programs along with extended
material are available on the website~\cite{mezzo}.

The soundness of \mezzo has been machine-checked~\cite{mezzo-proof}. In the
future, we wish to extend the language and its soundness proof with concurrency,
to guarantee data-race freeness.



\bibliography{english,local,icfp}

\begin{thebibliography}{1}

\bibitem{morrisett-al-07}
Amal Ahmed, Matthew Fluet, and Greg Morrisett.
\newblock ${L}^3$: {A} linear language with locations.
\newblock {\em Fundamenta Informaticæ}, 77(4):397--449, 2007.

\bibitem{bierhoff-aldrich-07}
Kevin Bierhoff and Jonathan Aldrich.
\newblock Modular typestate checking of aliased objects.
\newblock In {\em Object-Oriented Programming, Systems, Languages, and
  Applications (OOPSLA)}, pages 301--320, 2007.

\bibitem{boyland-nesting-10}
John~Tang Boyland.
\newblock Semantics of fractional permissions with nesting.
\newblock {\em ACM Transactions on Programming Languages and Systems}, 32(6),
  2010.

\bibitem{ohearn-07}
Peter~W. O'Hearn.
\newblock Resources, concurrency and local reasoning.
\newblock {\em Theoretical Computer Science}, 375(1--3):271--307, 2007.

\bibitem{mezzo-proof}
François Pottier.
\newblock Type soundness for {Core Mezzo}.
\newblock Unpublished, January 2013.

\bibitem{mezzo}
François Pottier and Jonathan Protzenko.
\newblock \mezzo.
\newblock \url{http://gallium.inria.fr/~protzenk/mezzo-lang/}, January 2013.

\bibitem{icfp-long}
François Pottier and Jonathan Protzenko.
\newblock Programming with permissions in \mezzo (long version).
\newblock Unpublished, March 2013.

\bibitem{reynolds-02}
John~C. Reynolds.
\newblock Separation logic: {A} logic for shared mutable data structures.
\newblock In {\em Logic in Computer Science (LICS)}, pages 55--74, 2002.

\bibitem{alias-types-00}
Frederick Smith, David Walker, and Greg Morrisett.
\newblock Alias types.
\newblock In {\em European Symposium on Programming (ESOP)}, volume 1782 of
  {\em Lecture Notes in Computer Science}, pages 366--381. Springer, 2000.

\end{thebibliography}


\end{document}